# The Effect of Hybrid Photovoltaic Thermal Device Operating Conditions on Intrinsic Layer Thickness Optimization of Hydrogenated Amorphous Silicon Solar Cells

M.J.M Pathak[1], K. Girotra[2], S.J. Harrison[1] and J.M. Pearce[1,3*]

1. Department of Mechanical and Materials Engineering, Queen's University, Kingston, ON, Canada

2. ThinSilicon Corporation, Mountain View, CA, USA

3. Department of Materials Science & Engineering and the Department of Electrical & Computer Engineering, Michigan Technological University, USA

* Contact author
Joshua M. Pearce
Department of Materials Science & Engineering
Department of Electrical & Computer Engineering
Michigan Technological University
601 M&M Building
1400 Townsend Drive
Houghton, MI 49931-1295
906-487-1466
pearce@mtu.edu

**Abstract**

Historically, the design of hybrid solar photovoltaic thermal (PVT) systems has focused on cooling crystalline silicon (c-Si)-based photovoltaic (PV) devices to avoid temperature-related losses. This approach neglects the associated performance losses in the thermal system and leads to a decrease in the overall exergy of the system. Consequently, this paper explores the use of hydrogenated amorphous silicon (a-Si:H) as an absorber material for PVT in an effort to maintain higher and more favourable operating temperatures for the thermal system. Amorphous silicon not only has a smaller temperature coefficient than c-Si, but also can display improved PV performance over extended periods of higher temperatures by annealing out defect states from the Staebler-Wronski effect. In order to determine the potential improvements in a-Si:H PV performance associated with increased thicknesses of the i-layers made possible by higher operating temperatures, a-Si:H PV cells were tested under 1 sun illumination (AM1.5) at temperatures of 25°C (STC), 50°C (representative PV operating conditions), and 90 °C (representative PVT operating conditions). PV cells with an i-layer thicknesses of 420, 630 and 840 nm were evaluated at each temperature. Results show that operating a-Si:H-based PV at 90 °C, with thicker i-layers than the cells currently used in commercial production, provided a greater power output compared to the thinner cells operating at either PV or PVT operating temperatures. These results indicate that incorporating a-Si:H as the absorber material in a PVT system can improve the thermal performance, while simultaneously improving the electrical performance of a-Si:H-based PV.

## 1. Introduction

Thin-film-based hydrogenated amorphous silicon (a-Si:H) solar photovoltaic (PV) cells hold promise over conventional crystalline silicon (c-Si) PV due to technical attributes including a higher absorption coefficient, less material requirements, excellent ecological balance sheet, low manufacturing costs and a high return on investment in manufacturing [1-3]. This promise, however, is limited by the Staebler-Wronski effect (SWE), which causes a-Si:H cell performance to degrade when exposed to sunlight [4-9]. SWE can be minimized with the use of optical enhancement and by thinning the intrinsic layer (i-layer), but the consequence is a decrease in initial performance [1, 10]. It is well established that annealing SWE defects can be achieved, not only at elevated temperatures over extended time periods in the dark (e.g. 150 °C for 4 hours), but also at room temperature [4, 11-13]. Furthermore, it is known that SWE is limited at higher temperatures [8], but this effect cannot be utilized at standard operating temperatures of PV because conventional modules cannot consistently achieve temperatures greater than 50 °C in ambient conditions in non-concentrated applications [14]. Thus despite more than 40 years of research in a-Si:H-based PV, the highest performing cells are fabricated sub-optimally thin to avoid unacceptable losses from SWE. Where real-estate with solar access is limited, the relatively low performance of a-Si:H-based PV also has limited (although large) markets.

Due to this challenge of limited optimal rooftop real-estate, there is a renewed interest in photovoltaic solar thermal (PVT) hybrid systems [15-18]. The typical PVT system is fabricated by simply adhering a standard c-Si-based PV module to a thermal system, which is far from optimal [19]. Historically, the main objective in PVT systems using c-Si PV is to cool the solar cell to achieve higher electrical outputs [20] because c-Si cells suffer performance losses of 0.4%/ °C [21]. However, as a result, the thermal efficiency of the system is sacrificed to meet the cooling demands [22]. A superior PVT design approach would incorporate both the thermal and electrical output in one optimized design.

Utilizing a-Si:H materials in replacement of c-Si in PVT systems offers a promising solution to this problem [23]. Firstly, a-Si:H can be deposited directly onto the absorber plate making an integral system [24]. Secondly, a-Si:H has a superior temperature coefficient (0.1%/ °C) [21] to c-Si-based PV. Lastly, running the system at optimal thermal conditions would allow for the a-Si:H cells to achieve a higher operating temperature [25], which would have the added benefit of annealing SWE defects and achieving a higher electrical performance [26].

In this paper a-Si:H is investigated as a potential absorber material in a PVT system in order to determine the potential improvements in a-Si:H PV performance with increased thicknesses of i-layers at PVT operating conditions. To accomplish this, a thickness series of a-Si:H solar cells with i-layers of 420, 630 and 840 nm were degraded under simulated solar irradiance (AM1.5, 1 sun). This degradation was repeated for a temperature series of 25 °C (Standard testing conditions [STC]), 50 °C (representative operating temperature of PV) and 90 °C (representative operating temperature of PVT). The kinetics of the degradation of the cells are presented and the implications of using a-Si:H for PVT applications is discussed.

## 2. Materials and Methods

Thin film a-Si:H PV cells were fabricated in a series of three i-layer thicknesses of 420 nm, 630 nm, and 840 nm on Asahi TCO U-type glass substrates. The cells were prepared in an AKT plasma enhanced chemical vapour deposition system. Figure 1 shows the device configuration used in this investigation from the bottom of the cell up including: AGC float glass (3 mm)/SnO$_2$:F (700 nm)/ Ag (200 nm)/ AZO (100 nm)/ n-a-Si:H 25 nm/ i-a-Si:H 420 nm to 840 nm/ p-a-Si:H 15nm/ ITO 70nm. PV cells were fabricated by ThinSilicon.

*{Insert Fig. 1}*

Samples were placed on a Chemat Technology Inc. hot plate, model KW-4AK, and secured with thermal paste. A k-type thermocouple was placed on the surface of the sample beside the cell of interest. The thermocouple readings were measured by a Cole Parmer Digi-Sense temperature controller k-type sensor, which was also used to control the temperature of the cell. A 12 V, 4.5 A computer fan was used to cool the cell for the 25 °C test. A PV Measurements class AAA solar simulator was used to irradiate the samples at 1 sun (AM 1.5 spectra) and K2400 I-V software was used with a Keithley 2400 source-meter and Keithley 2000 multi-meter to measure the current-voltage (I-V) output of the cells. An AutoIt macro was created to take measurements at 7 second intervals for the first hour, 30 seconds for the next 2 hours and then every 60 seconds for the remaining degradation time. The cells were degraded at the constant temperatures of 25, 50 and 90 °C with constant irradiance until the cell had stabilized. Light I-V measurements were made at the degradation temperature and a W/°C calibration test was made with the DSS samples.

## 3. Results and Discussion

Figures 2 to 4 show the a-Si:H-based PV device maximum power as a function of light soaking time at AM 1.5 and 1 sun. The temperature of the cells are measured at the degradation temperature. The effect of temperature on the power is -0.016 W/°C, which is similar to the results found by Sandia National Laboratories [27]. The kinetics of degradation show similar kinetics to those reported previously [28-31]. At 25 °C (STC) the thinner (420 nm) cells provide superior degraded steady-state (DSS) performance than thicker (630 nm and 840 nm) cells due to the SWE. Figure 2 shows that the maximum power of cells with i-layers of thickness of 420 nm degraded at 25 °C relative to the 630nm and 840nm cells degraded at 50 °C. As can be seen in Figure 2, if the thicker cells are degraded at higher temperatures they stabilize at a higher power than the thinner cells degraded at STC.

In Figures 2 and 3, the comparison between cells operating at different temperatures is shown. If the effect of temperature on the maximum power performance was incorporated into the comparison, the power output of the higher temperature thicker cells would greatly exceed the thinner cells at lower temperatures. This implies the when choosing the correct thickness for the a-Si:H in a PVT system, the 25°C DSS from SWE is not the limiting factor and that the operating temperature of the module should also be considered.

*{Insert Fig. 2}*

As has been suggested before, Figure 2 clearly shows that the use of STC in determining the design of an a-Si:H based PV cell is suboptimal [32]. If the thinner cells at operating temperatures (50 °C) are compared to thicker cells at higher temperatures (90 °C), the pattern is still the same for the higher temperatures as can be seen in Figure 3.

*{Insert Fig. 3}*

Even when the cells are compared at the same higher temperature of 90 °C, as seen in Figure 4, the thicker cells have superior performance. The power and energy gain from the 630nm over the 420nm is 1%. This demonstrates that when designing the PVT system, depending on the optimal absorber temperature, the thickness of the a-Si:H can be modified to produce the maximum electrical and thermal output.

*{Insert Fig. 4}*

Based on these findings, if the system is running at less than 90 °C, the thinner cells would be better suited for PVT applications and the opposite is true at temperatures greater than or equal to 90 °C. When running the system at 90 °C with the thicker cells, 630nm and 840nm, 2% and 0.5% more energy respectively, is produced than when operating with a thinner cell at the same temperature. The use of a-Si:H as the absorber layer in a PVT device also enables the option of utilizing short pulse thermal annealing cycles to better optimize the overall solar energy conversion efficiency [33]. During an annealing pulse solar energy is lost because the heat that would normally be collected is used to anneal the PV and the PV operating temperature is very high (e.g. 100°C). However, it was found when comparing the cells after stabilization at normal 50°C degradation that a pulse annealing sequence resulted in more than 10% electrical energy gain [33]. Further refinement of the promising preliminary work in this area is needed to generate a dispatch strategy for pulse thermal annealing cycles based on geographic location. Furthermore, based on the results of the work presented in this paper, future work is needed to design a system with an i-layer thickness that corresponds to the requirements of the thermal system. Future research is required to determine the most favourable thickness at any given operating temperature as the location of the system will dictate the desirable absorber temperatures. Finally, to optimize a-Si:H-based PVT systems these two effects must be taken into account (pulse annealing and thickness optimization) in order to balance the tradeoffs between PV cell design and dispatch strategy of annealing cycles in the PVT.

Published as: M.J.M Pathak, K. Girotra, S.J. Harrison and J.M. Pearce, "The Effect of Hybrid Photovoltaic Thermal Device Operating Conditions on Intrinsic Layer Thickness Optimization of Hydrogenated Amorphous Silicon Solar Cells" *Solar Energy* **86**, pp. 2673-2677 (2012). DOI: http://dx.doi.org/10.1016/j.solener.2012.06.002## 4. Conclusions

The performance of a-Si:H-based solar photovoltaic devices are highly dependent on operating temperature. The higher the temperature (up to 90 °C) the higher the DSS is at a given thickness, due to annealing of SWE defects. This study found that this annealing effect could be utilized for targeting a PVT absorber material with thicker a-Si:H PV layer. When comparing the thicker cells running at 90 °C, to that of the thinner cells at 50 °C, it was found that the thicker cells produced 1% more power and energy under 1 sun illumination. Running all the cells at 90 °C, the thicker cells performed better than the thinner cells by 0.5% and 2% for the 840nm and 630nm. These results imply that utilizing a thicker intrinsic layer of a-Si:H for use in PVT systems will allows for the system to be optimized both for the electrical and thermal aspects.

### Acknowledgements

This work was supported by the Natural Sciences and Engineering Research Council of Canada, Canada Foundation for Innovation, Ministry of Research and Innovation, the Canadian Solar Building Network, and PV Measurements Inc.### References

[1] Rech, B., Wagner, H., 1999. Potential of amorphous silicon for solar cells. Appl. Phys. A. 69, 155–167.
[2] Pearce, J.M., 2002. Photovoltaics – a path to sustainable futures. Futures. 34, 663-674.
[3] Branker, K., Pearce, J.M., 2010.Financial return for government support of large-scale thin-film solar photovoltaic manufacturing in Canada. Energ. Policy. 38, 4291–4303.
[4] Staebler, D.L., Wronski, C.R., 1977. Reversible conductivity changes in discharge- produced amorphous Si. Appl. Phys. Lett. 31, 292–294.
[5] Fritzsche, H., 2001. Development in understanding and controlling the Staebler-Wronski Effect in a-Si:H. N. Rev. Mater. Res. 31, 47-79.
[6] Wronski, C.R., Pearce, J.M., Deng, J., Vlahos, V., Collins, R.W., 2004. Intrinsic and light induced gap states in a-Si:H materials and solar cells – effects of microstructure. Thin Solid Films. 451-452, 470-475.
[7] Deng, J., Albert, M.L., Pearce, J.M., Collins R.W., Wronski, C.R., 2005. The nature of native and light induced defect states in i-layers of high quality a-Si:H solar cells derived from dark forward bias current-voltage characteristics. Mater. Res. Soc. Symp. Proc. 862, A11.4.
[8] Ruther, R., T-Mani, G., del Cueto, J., Adelstein, J.,. Dacoregio, M.M, von Roedern, B., 2005. Performance test of amorphous silicon modules in different climates - year three: higher minimum operating temperatures lead to higher performance levels. 31st IEEE Photovoltaic Specialists Conf. Proc. 1635-1638.
[9] Klaver, A., van Swaaij, R., 2008. Modeling of light-induced degradation of amorphous silicon solar cells. Sol Energ. Mat. Sol. C. 92, 50-60.
[10] Alkaya, R., Kaplan, H., Canbolat, S.S., Hegedus, A., 2009. Comparison of fill factor and recombination losses in amorphous silicon solar cells on ZnO and SnO2. Renew. Energ. 34, 1595-1599.
[11] Fujikake, S., Ota, H., Ohsawa, M., Hama, T., Ichikawa, Y., Sakai, H., 1994. Light-induced recovery of a-Si solar cells. Sol. Energ. Mat. Sol. C. 34, 449-454.
[12] Albert, M.L., Deng, J., Niu, X., Pearce, J.M., Collins, R.W., Wronski, C.R., 2005. The creation and relaxation kinetics of light induced defects in a-Si:H located at different energies in the gap. Mater. Res. Soc. Symp. Proc. 862, A13.2.
[13] Pearce, J.M, Deng, J., Albert, M.L., Wronski, C.R., Collins, R.W., 2005. Room temperature annealing of fast states from 1 sun illumination in protocrystalline Si:H materials and solar cells. 31st IEEE Photovoltaic Specialists Conf. Proc. 1536-1539.

**Figure Captions**

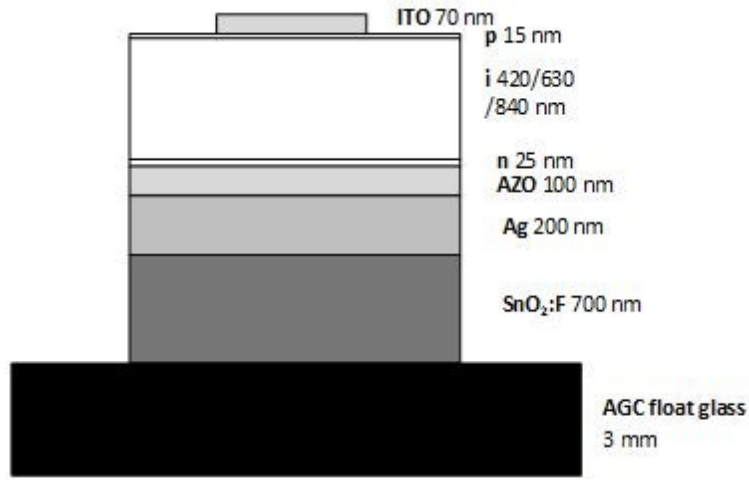

Figure 1: Device schematic

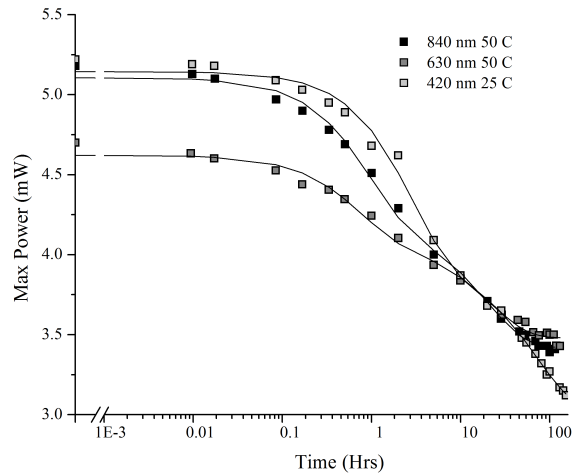

Figure 2: Comparison of the maximum power as a function of light soaking time for cells with i-layers of thickness of 420nm and degraded at 25 ºC to the 630nm and 840nm cells at 50 ºC

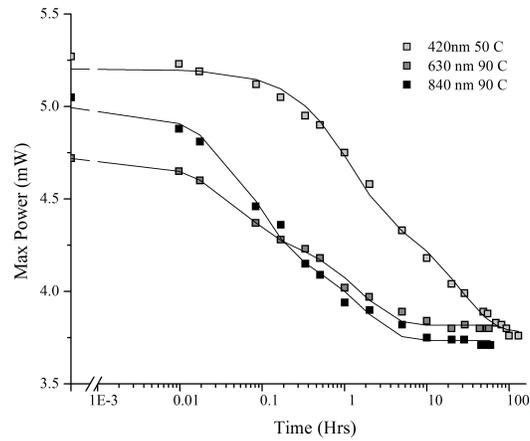

Figure 3: Comparison the maximum power as a function of light soaking time for cells with i-layers of thickness of 420nm cells and degraded at 50 ºC, to the 630 and 840 nm cells at 50 and 90 ºC

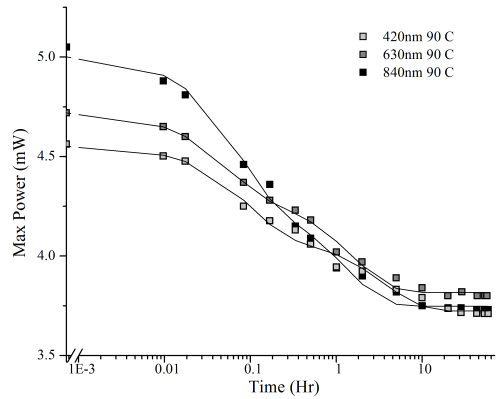

Figure 4: Comparison the maximum power as a function of the light soaking time of cells with i-layers of thickness of 420nm to 630 and 840nm at 90 ºC